# Understanding Crypto-Ransomware

In-Depth Analysis of the Most Popular Malware Families

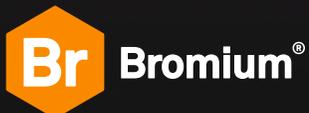

# Table of Contents

AUTHORS

Vadim Kotov
Mantej Singh Rajpal

# Executive Summary

A bully stuffing a student into a locker is apocryphal, but on the Internet the reality is far worse. An emerging cybersecurity threat can encrypt files, locking them from user access, until a ransom is paid.

This threat is called crypto-ransomware (ransomware) and includes at least a half-dozen variants, including CryptoLocker and CryptoWall. Ransomware shows no sign of abating since traditional detection-based protection, such as antivirus, has proven ineffective at preventing the attack. In fact, ransomware has been increasing in sophistication since it first appeared in September 2013, leveraging new attack vectors, incorporating advanced encryption algorithms and expanding the number of file types it targets.

Ransomware lacks the subtlety of more traditional Trojan attacks that seek to evade detection and steal sensitive information, such as credit card numbers and bank account credentials. Instead, ransomware immediately makes its presence known by encrypting files and demanding payment for the keys to unlock them. Victims of the attack may experience anxiety or disbelief, so they are likely to pay the ransom to end the incident, often without reporting the crime in order to avoid further embarrassment.

There have been reports of thousands of Internet users plagued by this attack (and likely thousands more that have gone unreported). Among the most popular variants of ransomware, it is estimated that CryptoLocker and CryptoWall have collected millions of dollars from its victims.

Similar to online black markets, the creators of ransomware have been using traffic anonymizers, such as TOR, and anonymous currencies, such as Bitcoin, to receive ransom payments from their victims without being traced. Encouraged by the financial "success" of these variants, malware authors have developed several families of ransomware recently.

In this report, Bromium Labs dissects nearly 30 samples of ransomware variants that have been encountered since September 2013, revealing a trend of increasing sophistication. Key highlights from the analysis include:

- Ransomware proliferates through new attack vectors, such as malvertising, employing anti-analysis and persistence techniques to ensure system compromise

- Advanced encryption algorithms, such as RSA and AES, prevent decryption without the key

- Ransomware has shifted its attention to the enterprise, targeting more than 230 file types (up 200 percent from 70 file types in 2013)

## Introduction

Crypto-ransomware is a type of malware that encrypts files on the victim machine using strong cryptography. After that it notifies the user that their files were encrypted and demands ransom for decryption (see Figure 1). The decryption key is stored on the attacker's server so victims cannot recover their files without paying the ransom.

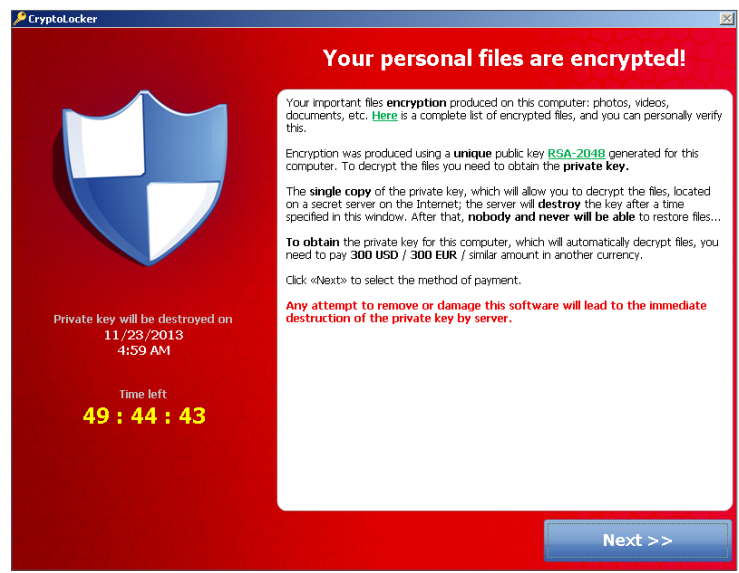

Figure 1: CryptoLocker ransom notification

Crypto-ransomware is rather different from traditional malware:

- It doesn't steal victim's information—instead it makes it impossible to access your information
- It doesn't try to remain stealthy after files are encrypted because detection won't restore the lost data
- It's relatively easy to produce—there are a number of well-documented crypto-libraries

Crypto-ransomware compromises the endpoint through one of the following attack vectors:

- Spam / Social engineering [1]
- Direct drive-by-download [2]
- Drive-by-download through malvertising [3]
- Malware installation tools and botnets [4]

This makes crypto-ransomware one the nastiest threats of the past year. Given that the actors behind these campaigns have collected millions of dollars [5], it seems like infected users continue to pay the ransom.

In this report, we analyze six families of crypto-ransomware that appeared during 2013-2014:

1. Dirty Decrypt
2. CryptoLocker
3. CryptoWall / Cryptodefense
4. Critroni / CTB Locker
5. TorrentLocker
6. Cryptographic Locker

We describe in detail our analysis methodology and then share our findings. The results are split into several categories including "Droppers", "Command-and-Control (C&C) communication", "Encryption" and "Targeted File Types."

Appendices contain source code of the scripts and programs we wrote for this research.

# Dataset and Timeline

Let's begin by establishing the chronological order of the samples analyzed. Figure 2 depicts a timeline of families based on their earliest sighting in the wild. These dates are based on various reports, coupled with our own analysis.

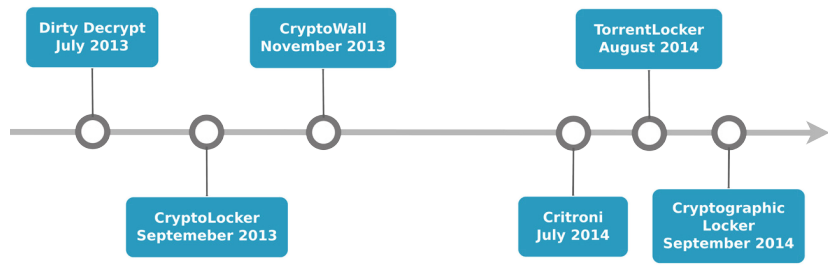

Figure 2: Approximate timeline for crypto-ransomware

CryptoLocker surfaced in the fall of 2013, and remained among the most widespread of the crypto-ransomware families until mid-2014 [6].

CryptoWall surfaced towards the end of 2013 [7], but didn't become prominent until 2014 [8]. New strains of CryptoWall have appeared as recently as last month (October 2014).

Critroni behaves similarly to CryptoWall—they both require using the TOR browser to make payments, and they both were prominent around the summer of 2014 [9,10].

Our sample of DirtyDecrypt outdates CryptoLocker, appearing in the summer of 2013—a few months before CryptoLocker became prominent [11]. This sample only targets and encrypts eight different file formats, which makes sense due to it being among the earliest iterations of ransomware.

Table 1 shows the malware samples we observed in this research.

TABLE 1: LIST OF MD5 HASHES OF THE SAMPLES ANALYZED WITH
THE CORRESPONDING COMPILATION DATES

CryptoLocker

| NO | MD5 | COMPILATION DATE |
|----|-----|------------------|
| 1 | d95bf36c4edf480fe9fd208e44c72be4 | 5/15/2014 |
| 2 | 04fb36199787f2e3e2135611a38321eb | 9/7/2013 |
| 3 | 180753f31b8295751aa3d5906a297511 | 9/11/2013 |
| 4 | 0204332754da5975b6947294b2d64c92 | 10/7/2013 |
| 5 | 2a1609ef72f07abc97092cb456998e43 | 12/9/2013 |
| 6 | 685634dac8b4c2b9429e80f8cd572563 | 1/20/2014 |
| 8 | 7f3cc059ffc6c11fe42695e5f19553ab | 12/3/2013 |
| 9 | 7f9c454a2e016e533e181d53eba113bc | 11/19/2013 |
| 10 | a8e0d4771c1f71709ddb63d9a75dc895 | 10/14/2013 |
| 11 | bbb445901d3ec280951ac12132afd87c | 10/21/2013 |
| 12 | e93af50428fcc74af931bfed7a1dc1b2 | 3/4/2014 |
| 13 | f1e2de2a9135138ef5b15093612dd813 | 5/8/2004* |
| 14 | 44217c15f30538a1fbdf614c9785c9b7 | 3/28/2011* |

Cryptowall/Cryptodefense

| NO | MD5 | COMPILATION DATE |
|----|-----|------------------|
| 15 | 73a9ab2ea9ec4eaf45bce88afc7ee87e | 8/15/2014 |
| 16 | 0650c9045814c652c2889d291f85c3ae | 6/2/2014 |
| 17 | b6c7943c056ace5911b95d36ff06e0e4 | 5/3/2014 |
| 18 | 90a0231b5d41c33bbe352bc3dab6b3a6 | 6/12/1979* |
| 19 | c1ea1ac134f5412af555e8b7ea8a8a54 | 5/29/2014 |
| 20 | e2e6674fc6ae6302ce8959b6686e1271 | 3/30/2010 |
| 21 | 31c2d25d7d0d0a175d4e59d0b3b2ec94 | 10/1/2014 |
| 22 | 1ef4264c5b802b4e83c82c87ffbc323d | 9/8/2014 |
| 23 | a9927372adb1bbab4d9feda4973b99bb | 2/8/2093* |

Critroni / CTB Locker

| NO | MD5 | COMPILATION DATE |
|----|-----|------------------|
| 24 | e89f09fdded777ceba6412d55ce9d3bc | 7/10/2014 |

Dirty Decrypt

| NO | MD5 | COMPILATION DATE |
|----|-----|------------------|
| 25 | 7a3c8d7f8b2b5bd26995dd33f4c1ee3c | 6/25/2013 |

TorrentLocker

| NO | MD5 | COMPILATION DATE |
|----|-----|------------------|
| 26 | 93cbe4ed3d46abe732a124a41e7147a2 | 9/25/2014 |
| 27 | e982953f4b15ad41dbccb13a09970214 | 9/25/2014 |

Cryptographic Locker

| NO | MD5 | COMPILATION DATE |
|----|-----|------------------|
| 28 | c32354ee13930113072fdba163dc8ca4 | 8/28/2014 |

*The compilation date of these samples was either tweaked or compiled on a machine with incorrect date

The conclusions made in the rest of this report are applied to this particular dataset, which are a subset of all crypto-ransomware samples. The results in our report are consistent with other industry reports and analyses.

# Analysis Methodology

In this analysis we leveraged dynamic methods and experimentation in a controlled environment as opposed to traditional debugging/disassembling approaches. We focused primarily on CryptLocker and CryptoWall since they comprised the majority of our samples, 14 and 9 respectively. This enabled us to cross-analyze how a particular family changes from sample to sample.

CryptoLocker and CryptoWall share some common characteristics:

• They fetch a public key from the C&C and only then perform the encryption
• Use WinCrypto for file encryption

Therefore, the best approach to this analysis is:

1. Reverse engineer the C&C protocol for one sample of each family

2. Set up a fake C&C server so that malware can execute in our controlled environment

3. Use API Monitor to look up which crypto functions were called

4. Compare the behavioral patterns observed in each experiment

Fake C&C code for Cryptolocker and Cryptowall can be found in Appendix A and B respectively. We bumped into one complication with Cryptolocker however. Figure 3 describes its communication protocol and can be described as follows:

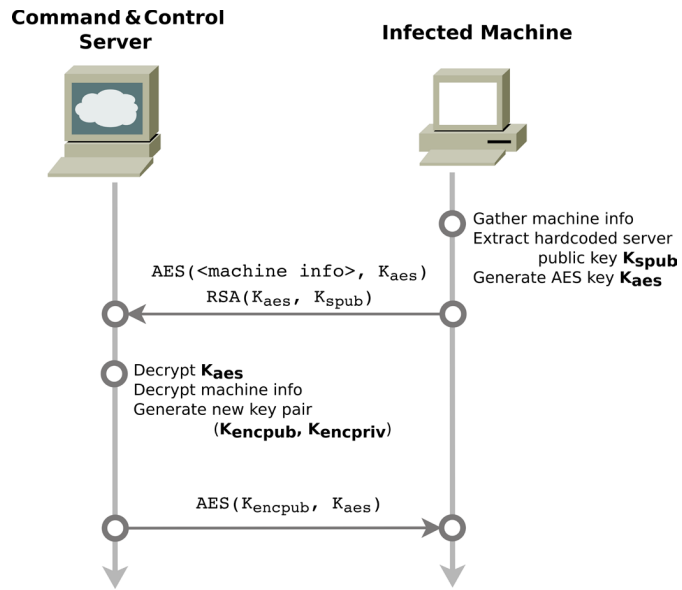

1. Client gathers victim machine info and encrypts it with an AES session key
2. The AES key is then encrypted using RSA and hardcoded C&C public key
3. Encrypted victim machine data and encrypted session key are concatenated and sent to the server
4. Server decrypts victim information and the session key using its private key
5. Server generates another key pair that is actually going to be used for file encryption
6. The public key is sent back to the client encrypted with the aforementioned session AES key

Figure 3: CryptoLocker C&C protocol

The problem is that in order to run CryptoLocker in our controlled environment, we need to possess the servers' private keys, which isn't feasible. The good news is that CryptoLocker uses WinCrypto, hence to decrypt the server message it calls CryptDecrypt. Let's look at the prototype:

```
BOOL WINAPI CryptDecrypt(

    _In_      HCRYPTKEY hKey,

    _In_      HCRYPTHASH hHash,

    _In_      BOOL Final,

    _In_      DWORD dwFlags,

    _Inout_   BYTE *pbData,

    _Inout_   DWORD *pdwDataLen

);
```

CryptDecrypt takes a pointer to the ciphertext (*pbData*) and its length, and as a result it replaces the data pointed at by *pbData* with the decrypted message. Subsequently, *pdwDataLen is set* to the length of the plaintext. Alternately: CryptDecrypt can easily be hooked and bypassed without interfering with the program workflow. This allows us to communicate with the client without possessing the respective private key. Source code and setup instructions for hooking can be found in Appendix A.

Before we move on to discussing the results, one final remark should be made. The vast majority of the samples we dealt with were obfuscated, and several of them detected our debugger (even with a number of anti-debugging plugins for OllyDbg). Instead of individually reversing each anti-debugging trick, we used another approach. We noticed that typical self-de-obfuscation performed by the malware analyzed consists of the following steps:

1. Create another instance of itself (or a process to inject the code into) in suspended state

2. Unmap the executable image from the target process

3. Unpack the payload

4. Write and map the payload to the target process

5. Resume the process

This method uses the *WriteProcessMemory* API call:

```
BOOL WINAPI WriteProcessMemory(
    _In_    HANDLE hProcess,
    _In_    LPVOID lpBaseAddress,
    _In_    LPCVOID lpBuffer,
    _In_    SIZE_T nSize,
    _Out_   SIZE_T *lpNumberOfBytesWritten
);
```

So what we could do is hook this call and dump the buffer for size *nSize* pointed by *lpBuffer*. We also need to save the address where the data is written (*lpBaseAddress*) in order to restore the actual image written. The source code of the hooking program can be found in Appendix C.

We considered dumping the target process after the code was injected, but we often got either a corrupted image or too many artifacts of the target process. Our approach, though not perfect, is simple to implement and quite stable.

## Results

### Droppers, anti-analysis and persistence

**Analysis summary:**

- All the samples have fairly complex obfuscation and use covert launch mechanisms
- Crypto-ransomware generates a number of easily detectable indicators of compromise. After files were encrypted stealthiness is no longer a priority

**Details:**

The most common mode of operation for crypto-ransomware droppers is process injection. It is done by creating a suspended process (such as explorer.exe or svchost.exe) and swapping the image with the unpacked payload. Sometimes this becomes pretty complex, involving several layers of de-obfuscation and process injection, such as the CryptoWall incidents we analyzed (Figure 4).

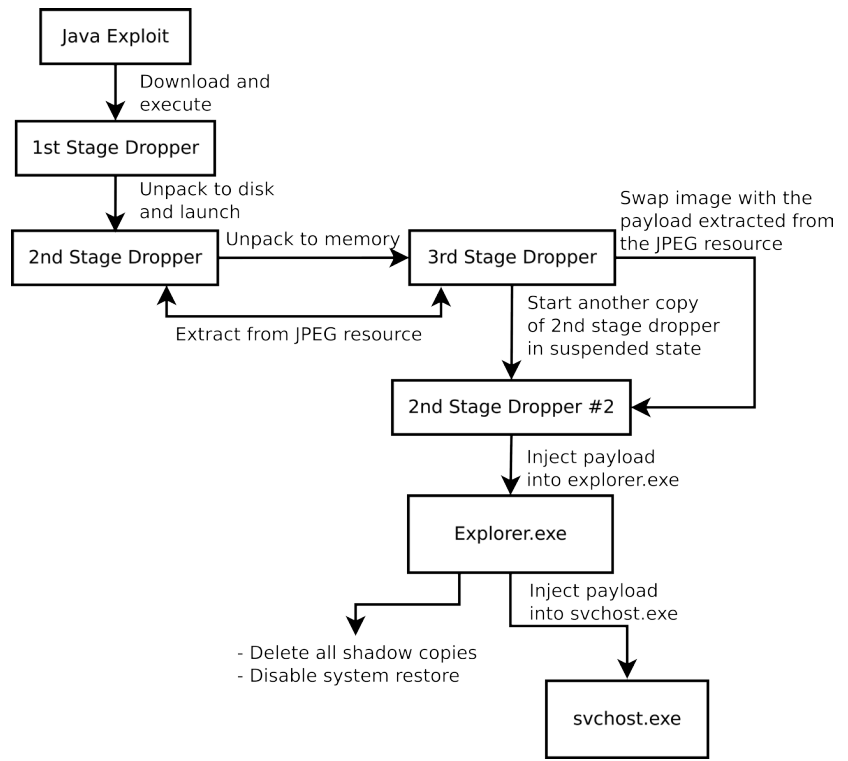

Figure 4: CryptoWall infection workflow

A number of samples will terminate early if run under the debugger, which suggests that crypto-ransomware developers (or whoever they buy droppers from) are implementing techniques to interfere with detection and analysis.

All the samples analyzed use registry and mutexes that are easily detectable indicators of compromise. This means that they don't care about being detected because it doesn't matter once the victim's files are encrypted.

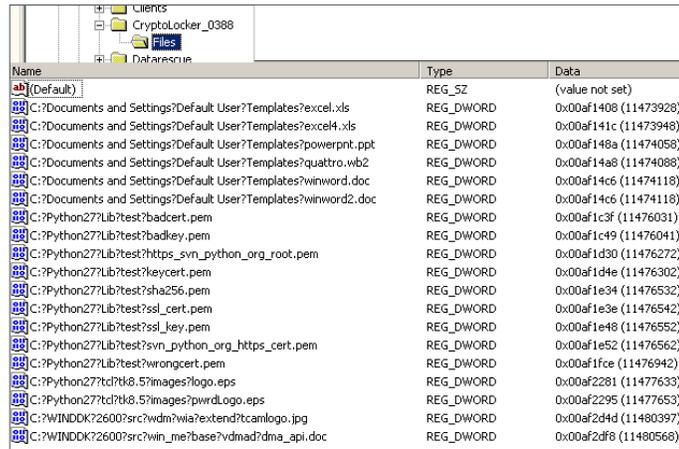

Figure 5: CryptoLocker keeps the list of encrypted files in the registry

CryptoLocker stores the list of files it encrypts (as shown on Figure 5). In this instance, the name was "Cryptolocker_0388", so an antivirus could detect the locker, but deleting this key would prevent a user from looking up which files were actually encrypted.

## C&C communication

### Analysis summary:

- Communication protocols have evolved from plaintext (HTTP) to encrypted (TOR, HTTPS)

- As a result C&C domains changed from those based on a domain name generator algorithm (DNGA) to hardcoded URLs since encrypted communication is harder to track during efforts to take down servers

- Early samples would encrypt files after contacting C&C, which could enable security teams to prevent infection by monitoring the traffic and terminating the connection before the malware could finish its job. Critroni / CTB Locker authors addressed this situation by encrypting files first and then communicating to C&C

Details:

Table 2 shows communication protocols and types of URLs in crypto-ransomware analyzed. The families are listed in chronological order.

TABLE 2: COMMUNICATION PROTOCOLS AND HOW C&C DOMAIN NAMES ARE STORED

| FAMILY | PROTOCOL | C&C DOMAINS |
|---|---|---|
| Dirty Decrypt | HTTP | DNGA |
| CryptoLocker | HTTP | DNGA and hardcoded URLs |
| CryptoWall / CryptoDefense | HTTP and later TOR | Hardcoded URLs |
| Critroni / CTB Locker | TOR | Hardcoded URLs |
| TorrentLocker | HTTPS | Hardcoded URLs |
| Cryptographic Locker | HTTP | No-IP / No-DNS, hardcoded |

Based on our observations, it seems crypto-ransomware developers have switched from plaintext protocols to protected communication using TOR and SSL. Although HTTP data is usually encrypted (e.g., CryptoLocker uses AES+RSA to encrypt POST request body) it still can be fingerprinted for detection purposes. Using completely encrypted channels such as TOR or HTTPS makes it impossible to write a network signature for early detection. For example, consider a CryptoWall request (Figure 6).

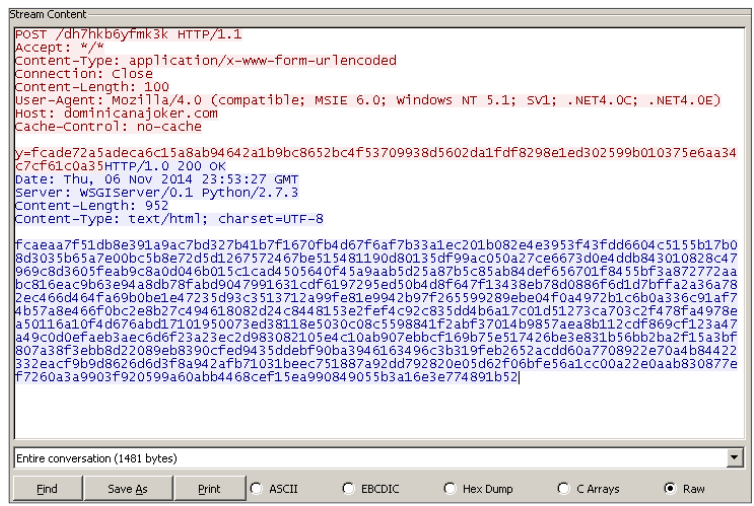

Figure 6 : CryptoWall request to C&C

Although the string sent is encrypted with RC4 it can be easily fingerprinted. The POST body can be described by the regular expression `[a-z]\=[a-z0-9]+` and for a deeper analysis it can even be decrypted since the path of the requested Web resource is an RC4 key. So when communication is encrypted there's no way of knowing which application creates the packets.

TOR might not be the best option either because it might be forbidden in certain networks or TOR related traffic might trigger IDS/IPS alerts. TorrentLocker's approach is the most stealthy because it is indistinguishable from the legitimate SSH connections made by browsers.

Another important aspect of crypto-ransomware workflow is when exactly the files get encrypted. Earlier samples such as CryptoLocker and CryptoWall first contact the server and only then perform encryption.

### Encryption

Analysis summary:

- Over time, encryption strength tends to increase, from RC4 to RSA+AES to ECDH+AES
- Apart from minor flaws, the cryptography is strong and implemented by the book (except for Dirty Decrypt and TorrentLocker)
- Crypto implementation evolves from using WinCrypto (easy to hook and dump private key) to statically linked OpenSSL code (more challenging to apply behavioral detection to)

Details:

The malware families analyzed use the following encryption schemes:

| FAMILY | ENCRYPTION ALGORITHMS | IMPLEMENTATION |
| --- | --- | --- |
| Dirty Decrypt | RC4 to encrypt whole files, RSA to then encrypt first 1024 bytes of each file | Inline |
| CryptoLocker | AES for file encryption, RSA for AES key encryption | MS Crypto API |
| CryptoWall | RSA for file encryption | MS Crypto API |
| Critroni / CTB Locker | AES for file encryption, ECDH for AES key encryption | OpenSSL, statically linked |
| TorrentLocker | AES for file encryption** | Allegedly OpenSSL,* statically linked |
| Cryptographic Locker | AES for file encryption | MS Crypto API (.NET) |

 * We couldn't reliably determine which library was used
** Although AES constants were found in the code it doesn't seem to use it (see Flaws and Version Evolution
   for more details)

Apart from several flaws found in early samples of CryptoWall [12] and TorrentLocker [13], the cryptography appears to be implemented by the book. The only exception is CryptoWall since encrypting whole files with RSA is in theory considered insecure (since RSA is a completely deterministic algorithm). It is worth noting this process is quite intense on memory and CPU, which might be used as a behavioral detection indicator. In other cases block or stream ciphers were used.

One question is why Critroni developers used elliptic curve cryptography to encrypt symmetric keys? In theory, it keeps keys short and remains as strong as RSA, as well as performing more efficiently, but it doesn't make too big a difference.

Using RSA still makes it impossible to restore AES keys without the private key, so why bother? Possible answers could be:

• Developers are afraid of making mistakes in implementing RSA since there are a number of attacks on this cipher [14]. In fact, a tiny mistake in implementation can lead to compromising the secrecy of encrypted data. Therefore, the authors chose to go with a less established and less analyzed approach based on elliptic curves. Of course, ECC might be as prone to errors as RSA, but it has fewer known attacks and analysis papers.

- Or perhaps this is simply a marketing move for adverts on underground message boards. In this case, using elliptic curve cryptography is meant to attract customers since it is a fancy feature that other crypto-ransomware does not have.

A primary strength of crypto-ransomware is its ability to use well-known and reputable crypto libraries to perform encryption. Interestingly, early families such as CryptoLocker and CryptoWall relied on Microsoft CryptoAPI, which may be considered a drawback since it is trivial to hook encryption routines. That makes early detection easier and allows intercepting session keys. Other families switched to statically linking the encryption functions to address this problem.

### Targeted File Types

Analysis summary:

- The number and type of targeted files continues to grow
- Attackers carefully select which files to encrypt—targeted files are not random
- Latest crypto-ransomware families aim at enterprises—they look for databases, CAD files and financial data

Details:

With the first three ransomware families coming to light in 2013, and the latter three in 2014, we can see a somewhat steady increase in the number of file formats being targeted (Figure 7).

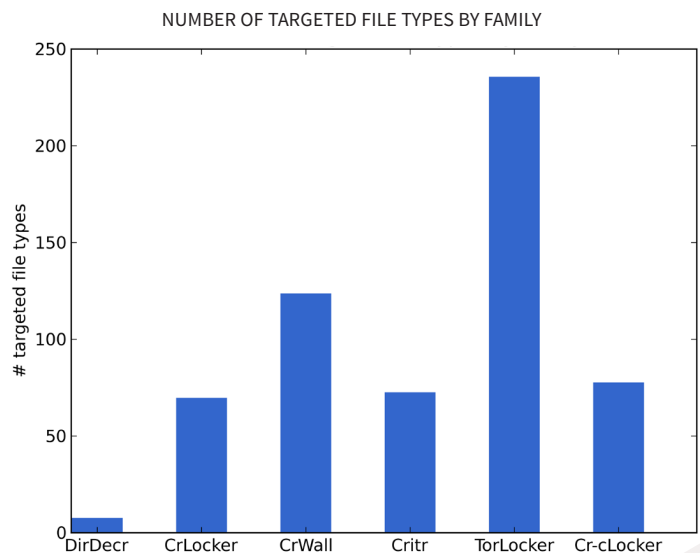

NUMBER OF TARGETED FILE TYPES BY FAMILY

Initially, the file types being targeted were more or less limited to a handful of text files and spreadsheets. Along came the game changer, CryptoLocker, encrypting an exponentially increased number of file types. From this point on, targeting more than 70 different extensions became a norm with crypto-ransomware. All types of music, videos and source code are generally among those encrypted.

TorrentLocker is a unique beast, targeting more than 200 types of files. Several obscure extensions that are not commonly used, such as *.djvu*, *.ycbcra* and *.blend* are among those targeted.

Let's look into targeted file types in more detail. First, let's define the list of categories:

- *doc*—all sorts of documents including text, word processor files, spreadsheets, etc.
- *img*—all images
- *av*—audio and video files
- *src*—source code files
- *cad*—all the possible design files
- *db*—databases
- *sec*—security related files including certificates, key chains and password managers
- *arch*—archives
- *fin*—all financial software from bank clients to accounts tools
- *bak*—various backups
- *oth*—formats that we were not able to determine or too rare to have its own category

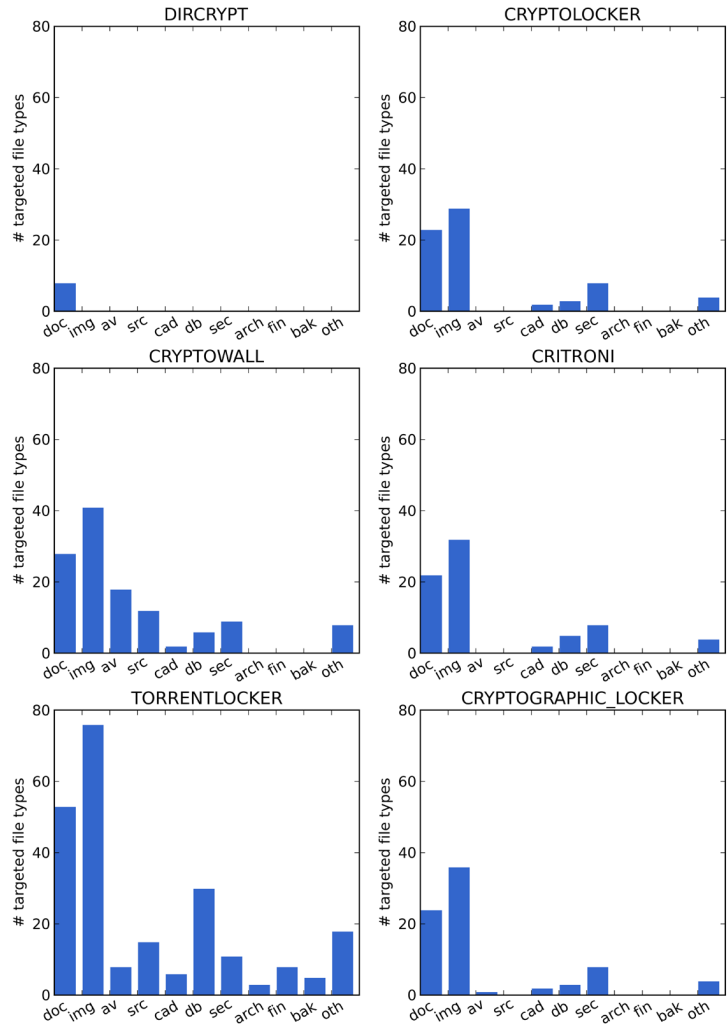

Figure 8: Comparison of targeted file categories

Figure 8 shows targeted categories of files and number of targeted file types per category. From this picture we can conclude that:

• Later families tend to target more file categories and formats

• Most targeted categories are documents and images

• Critroni's and Cryptographic Locker's plots are almost identical

• TorrentLocker is targeting far more file formats then other families

Interestingly crypto-ransomware looks for as many as 70 image formats. This includes all the popular extensions (such as PNG and JPG) and various types of raw images taken from professional cameras using the specialized software. For example '.ndd'—images from Nikon cameras or '.craw'—a common image format for several brands including Sony. Our analysis indicates that images are the most targeted file formats because the malware authors may realize the sentimental value of personal moments.

We suspect that malware developers didn't just dump all the known file extensions but actually carefully selected them. That's why we don't see too much of an increase in the *oth* (other) category.

Finally, earlier samples only targeted categories of files that can usually be found on an average user computer—documents and images for the most part. But later families (and especially TorrentLocker) really push on CAD and financial software. This means ransomware is aiming to infect enterprises. This makes sense since a home user losing data might not be such a disaster, while a company in the same situation can incur much more risk.

## Payment Options

### Analysis summary:

- Bitcoin is today's de facto standard for ransomware related transactions
- There is no clear trend in ransom price
- Later families set the price in BTC as opposed to actual currency

### Details:

It's interesting to note that the earliest CL samples offered 3-4 payment options, but as time progressed, Bitcoin became the preferred method of payment (Table 3).

TABLE 3: PAYMENT OPTIONS AND PRICES

| FAMILY | PAYMENT OPTIONS | PRICE OF DECRYPTING |
|---|---|---|
| Dirty Decrypt | Must use a pre-paid method: PaySafeCard, MoneyPak, etc. | 100 USD |
| CryptoLocker | Started off with 4 options (Bitcoin, MoneyPak, UKash, CashU). Eventually, narrowed in on BitCoin (MoneyPak was an option for US as well, but more expensive) | 300 USD |
| CryptoWall | Typically, payments required Bitcoin. Similarly to CL, earlier variants allowed pre-paid cards | 500 USD—increased to 1000 USD if not paid within time frame [8] |
| Critroni | Payments must be made in Bitcoin | 0.5 USD [15] |
| TorrentLocker | Bitcoin | 0.8 BTC |
| Cryptographic Locker | Bitcoin | 0.2-0.5 BTC [16] |

It's not surprising that Bitcoin has unanimously become the payment option of choice for crypto-ransomware. First, it's important to note that Bitcoin transactions cannot be reversed; only refunded by the receiver of the funds [17]. By using Bitcoin, ransomware authors are assured that no revocations will occur. Secondly, the use of mixing services, which are commonly used for money laundering, can easily cause illegally obtained funds to appear legitimate [18]. These services will mix large amounts of money, obfuscating the trail back to the original source. This is equivalent to moving US dollars through a bank with very strict secrecy laws [18]. Though there are several Bitcoin mixing services readily available, creating your own mixing service isn't difficult, even for those without a programming background [19].

Another interesting observation is that the latest family of crypto-ransomware, that is Cryptographic Locker requests a comparatively small ransom given the trend and the hype of the threat. It also was the easiest to implement since it was written in .NET, so perhaps the amount of money requested corresponds to the amount of resources committed to development.

Later families request ransom in BTCs as opposed to actual money (sometimes they put the actual price but that's merely an estimation based on the current BTC rates). This makes the prices dependent on the BTC rate. An interesting consideration for the future, does (or will) crypto-ransomware affect BTC cost?

## Implementation, Flaws and Version Evolution

**Analysis summary:**

- Most samples were written in C/C++, aside from Cryptographic Locker that was written in .NET
- A later version of CryptoWall has 97% of Critroni code—indicating some cooperation between gangs (or it's the same gang)
- Early samples contained several flaws that allowed restoring (at least partially) encrypted files
- The flaws are usually fixed in the next version or next family, which means that malware authors also read security reports

**Details:**

Most families analyzed were written in C/C++ and compiled using MSVC except for Cryptographic Locker. This one is .NET application so it is much easier to analyze, but it shows how easy it is to write a piece of crypto-ransomware. Also some droppers were written in Visual Basic.

We checked code resemblance between various samples, but there is not any considerable similarity. There was one interesting observation. Late versions of CryptoWall use TOR for C&C communication. However the TOR component is fetched from the Web before actual command and control. This piece of binary code has 97% similarity to Critroni / CTB Locker with 99% confidence (according to BinDiff[1]). Clearly the code was reused. This means that either these two families have the same developer or the developers partnered together.

For the past year, several flaws were spotted in certain crypto-ransomware families by various security researchers.

One of the earliest samples, Dirty Decrypt, was completely broken by Checkpoint [20]. Apparently, authors of this malware did not expect people to dive into the encryption protocol to create a decryption algorithm.

CryptoLocker was much better designed and used strong cryptography by the book. However, its victims could use shadow copies or restore points to partially restore their lost files.

In response to this, Cryptowall developers added the following command line calls to their product:

- `vssadmin.exe Delete Shadows /All /Quiet`—to delete the shadow copies
- `bcdedit /set {default} recoveryenabled No`—to disable recovery
- `bcdedit /set {default} bootstatuspolicy ignoreallfailures`—to disable windows error recovery on startup

Plus it sets DisableSR registry key to 1, which disables system restore. On the systems with user account control this won't work without admin privileges.

They however also made a mistake. The key pair was generated on the victim machine and authors forgot to delete the private key [12]. Later an article appeared describing how to decrypt the files locked by CryptoWall [21]. That was fixed in later versions and now the public key is fetched from the C&C.

Finally one of the TorrentLocker samples has a weird ad-hoc encryption scheme [13] that is a form of a stream cipher. Furthermore it uses one key and thus it was also broken.

# Conclusion

By every metric, crypto-ransomware continues to become more complex and more dangerous. Secure communication and elusive infection workflows make it nearly impossible for traditional detection-based security solutions, such as antivirus, to prevent the attack before the file encryption while a huge number of targeted file types endanger both home users and enterprises. Let's summarize the major points made in this paper:

- A number of families appeared over the past 18 months. We analyzed six, but there are more (e.g., SynoLocker and ZeroLocker)

- Crypto-ransomware uses every possible attack vector to get into victim machines

- Samples analyzed use fairly complex obfuscation and covert launch techniques that allow them to evade detection in the early stages of infection

- Communication with command-and-control servers is encrypted and extremely hard to spot in network traffic

- Cryptography used in the samples analyzed is for the most part implemented correctly and encrypted files are impossible to recover without a key

- The latest families target huge number of file formats from documents and images to CAD files and financial data

- All recent ransomware accepts payment in Bitcoins only. Apparently there's a good way of laundering BTC or maybe even a service on the black market

- Crypto-ransomware matures and evolves from version to version, additional features are added to ensure that files are impossible to recover (e.g., deleting shadow copies) and flaws are getting fixed

Prevention of such a threat is possible only in early stages of infection before files are encrypted. Antivirus and HIPS have two windows of opportunity to prevent the attack:

- At stage of drive-by exploit
- At stage of process injection

After that the malware will proceed with file encryption and detecting it at this stage might be too late.

Here are some recommendations on how to minimize the losses in case of infection:

1. Regularly backup your data

2. Use an external hard drive for your backups. Unplug the drive after it's finished copying files

3. Always keep UAC enabled. A number of operations performed by crypto-ransomware require admin privileges

It is likely that we'll see more crypto-ransomware families and this threat won't go away anytime soon. The only way to make it go away is to stop paying thus rendering its business model unprofitable. But this unfortunately is much easier said than done.

# Appendix A: Fake CryptoLocker C&C Server and CryptDecrypt Hook

Dependencies:
• bottle (http://bottlepy.org/docs/dev/index.html)

```
from bottle import route, run, SimpleTemplate, static_
file, post, request, response

PORT = 80

@route('/')

@route('/<path:path>', method='ANY')

def index(path=None):

    return """1\x00172.16.10.10\x00-----BEGIN PUBLIC
KEY-----

MIIAAOCAQ8AMIIBCgKCAQEAx2zYo7MDPjA7KZnEiufT

A+/Xakry/rZBJU5dIrn/s9MUuCkcX5LXtz4XHdW+xbwUJR4

/3Mvk8NbU26T5CNPiIpJjDC7K6t1bO5ZcXGPtL6VwY61taxtBmyBw
qoDOOBTBCHHljz+fzAcvGrZjAZC4Vk+6i5JHjBwaG6dI4PxZFdR

AwIDAQA

-----END PUBLIC KEY-----

\x00"""

def main():

    run(host='', port=PORT)

if __name__ == '__main__':

    main()
```

Public key provided was generated and exported using OpenSSL:

```
$ openssl req -x509 -nodes -newkey rsa:2048 -keyout
privatekey.pem -out certificate.pem

$ openssl rsa -in privatekey.pem -pubout
```

Hooking code is based on Microsoft Detours 3.0 (http://research.microsoft.com/en-us/projects/detours/)

## hook_appinit.cpp of hook_appinit.dll:

```
#include <stdio.h>

#include <windows.h>

#include <wincrypt.h>

#include "detours.h"

static BOOL (WINAPI *Real_CryptDecrypt)(HCRYPTKEY,
HCRYPTHASH, BOOL, DWORD, BYTE *, DWORD *)=
CryptDecrypt;

BOOL WINAPI Hook_CryptDecrypt(HCRYPTKEY hKey,
HCRYPTHASH hHash, BOOL Final, DWORD dwFlags, BYTE
*pbData, DWORD *pdwDataLen){

    return TRUE;

}

BOOL APIENTRY DllMain(HMODULE hModule, DWORD dwReason,
LPVOID lpReserved){

    if (dwReason==DLL_PROCESS_ATTACH){

        DetourTransactionBegin();

        DetourUpdateThread(GetCurrentThread());

        DetourAttach(&(PVOID&)Real_CryptDecrypt, Hook_
CryptDecrypt);

        DetourTransactionCommit();

    } else if(dwReason==DLL_PROCESS_DETACH){

        DetourTransactionBegin();

        DetourUpdateThread(GetCurrentThread());

        DetourDetach(&(PVOID&)Real_CryptDecrypt, Hook_
CryptDecrypt);

        DetourTransactionCommit();

    }

    return TRUE;

}
```

To compile the hook DLL the following command should be executed (we used MSVC):

```
cl /LD hook_appinit.cpp advapi32.lib wincrypt32.lib detours.lib
```

To to install the hook DLL we use the following batch script:

```
move hook_appinit.dll C:\

reg add "HKEY_LOCAL_MACHINE\Software\Microsoft\Windows NT\CurrentVersion\Windows" /v AppInit_DLLs /t REG_SZ /d "C:\hook_appinit.dll"
```

Thus every process in the system will start with the hook DLL mapped in its memory space.

# Appendix B: Fake CryptoWall C&C Server

Dependencies:
• bottle (http://bottlepy.org/docs/dev/index.html)

```
from bottle import route, run, SimpleTemplate, static_
file, post, request, response

from array import array

PORT = 80

def rc4_ksa(key):

    keylen = len(key)

    S = range(256)

    j = 0

    for i in range(256):

        j = (j+S[i]+key[i%keylen])%256

        S[i], S[j] = S[j], S[i]

    return S

def rc4_prng_and_xor(ct, S_):

    S = list(S_)
```

```
    pt = []

    ctlen = len(ct)

    i = 0

    j = 0

    for c in ct:

    i = (i+1)%256

    j = (j+S[i])%256

    S[i], S[j] = S[j], S[i]

        k = (S[i]+S[j])%256

        pt.append(c^S[k])

    return pt
@route('/')

@route('/<path:path>', method='ANY')

def index(path=None):

    msg = """{216|1test.onion|1a2b|US|-----BEGIN PUBLIC
KEY-----

AQ8AMIIBCgKCAQEAx2zYo7MDPjA7KZnEiufT

A+/Xakry/rZBJU5dIrn/
s9MUuCkc9nn3DPl0AJ2a9AVny7DaO4bOmCHn2ma5qvyu

A7C7t/Sgd0N7oNuuWBPqy1JQrbTdBO9PdjAOWaYC+e/
hCX5LXtz4XHdW+xbwUJR4

AwI-----END PUBLIC KEY-----}"""

    key = path

    print 'Initial key:', key

    key_sorted = sorted(bytearray(key))

    print 'Sorted key:', array('B', key_sorted).
tostring()

    pname, pvalue = request.params.items()[0]

    ct = bytearray(pvalue.decode('hex'))

    S = rc4_ksa(key_sorted)

    pt = rc4_prng_and_xor(ct, S)

    print 'Client message:', array('B', pt).tostring()
```

```
    ct = rc4_prng_and_xor(bytearray(msg), S)

    msg = array('B', ct).tostring().encode('hex')

    print 'Our response:', msg

    return msg

def main():

    run(host='', port=PORT)

if __name__ == '__main__':

    main()
```

# Appendix C:
# Hooking WriteProcessMemory

**hook_wpm.cpp of hook_wpm.dll:**
The program simply writes the buffer into a file called <base address>.bin

```
#include <windows.h>

#include <stdio.h>

#include <wincrypt.h>

#include "detours.h"

static BOOL (WINAPI *Real_WriteProcessMemory)(HANDLE,
LPVOID, LPCVOID, SIZE_T, SIZE_T *)=WriteProcessMemory;

BOOL WINAPI Hook_WriteProcessMemory(HANDLE hProcess,
LPVOID lpBaseAddress, LPCVOID lpBuffer, SIZE_T nSize,
SIZE_T *lpNumberOfBytesWritten){

    HANDLE hFile;

    char filename[MAX_PATH];

    DWORD dwBytesWritten = 0;

    sprintf(filename, "%x.bin", lpBaseAddress);

    hFile = CreateFile(filename,

                    GENERIC_WRITE,

                    0,

                    NULL,

                    CREATE_NEW,
```

```
                            FILE_ATTRIBUTE_NORMAL,

                            NULL);
    if(hFile!=INVALID_HANDLE_VALUE){

        WriteFile(

                        hFile

lpBuffer,

                        nSize,

                        &dwBytesWritten,

                        NULL);

    }
    return Real_WriteProcessMemory(hProcess,
lpBaseAddress, lpBuffer, nSize, lpNumberOfBytesWritten);

}
extern "C" __declspec(dllexport) void DummyFunc(void){

    return;

}
BOOL APIENTRY DllMain(HMODULE hModule, DWORD dwReason,
LPVOID lpReserved){

    if (dwReason==DLL_PROCESS_ATTACH){

        DetourTransactionBegin();

        DetourUpdateThread(GetCurrentThread());

        DetourAttach(&(PVOID&)Real_WriteProcessMemory,
Hook_WriteProcessMemory);

        DetourTransactionCommit();

    } else if(dwReason==DLL_PROCESS_DETACH){

        DetourTransactionBegin();

        DetourUpdateThread(GetCurrentThread());

        DetourDetach(&(PVOID&)Real_WriteProcessMemory,
Hook_WriteProcessMemory);

        DetourTransactionCommit();

    }
    return TRUE;

}
```

In this case we didn't want our DLL to be injected into every process (although this would a viable strategy, but you're going to need to modify the source in order to preserve the name of the target process). Instead we used the following program to start an executable and push our DLL into its memory space:

**run.cpp of run.dll:**

```
#include <windows.h>
#include <stdio.h>
#include <string.h>
#include "detours.h"
int main(int argc, char **argv){
   STARTUPINFO si;
   PROCESS_INFORMATION pi;
   LPTSTR szCmdLine = NULL;
   CHAR szDllName[MAX_PATH];
   CHAR szDetouredDll[MAX_PATH];
   BOOL res;
   if (argc<3){
      printf("Usage: %s <DLL> <PROCESS [ARGS]>\n", argv[0]);
      return -1;
   }
   szCmdLine = GetCommandLine();
   res = DetourCreateProcessWithDllEx(
      argv[2],
      &szCmdLine[strlen(argv[0])+strlen(argv[1])+2],
      NULL,
      NULL,
      FALSE,
      0,
```

```
        NULL,

        NULL,

        &si,

        &pi,

        argv[1],

        NULL

        );

    return 0;

}
```

Bromium has transformed endpoint security with its revolutionary isolation technology to defeat cyber attacks. Unlike antivirus or other detection-based defenses, which can't stop modern attacks, Bromium uses micro-virtualization to keep users secure while delivering significant cost savings by reducing and even eliminating false alerts, urgent patching, and remediation—transforming the traditional security life cycle.

1   http://www.zynamics.com/bindiff.html

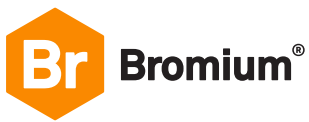